\documentclass[aps,showpacs,amsmath,amssymb,12pt]{revtex4}
\usepackage{latexsym}
\usepackage{bm}

\begin{document}

\title{Closed timelike curves and geodesics of G{\" o}del-type metrics}

\author{Reinaldo J. Gleiser}
\email{gleiser@fis.uncor.edu}
\affiliation{Facultad de Matem\'{a}tica, Astronom\'{i}a y F\'{i}sica,
Universidad Nacional de C\'{o}rdoba,\\ 
Ciudad Universitaria, (5000), C\'{o}rdoba, Argentina}

\author{Metin G{\" u}rses}
\email{gurses@fen.bilkent.edu.tr}
\affiliation{Department of Mathematics, Faculty of Sciences,\\
             Bilkent University, 06800, Ankara, Turkey}

\author{Atalay Karasu}
\email{karasu@metu.edu.tr}
\affiliation{Department of Physics, Faculty of Arts and  Sciences,\\
             Middle East Technical University, 06531, Ankara, Turkey}

\author{{\" O}zg{\" u}r Sar{\i}o\u{g}lu}  
\email{sarioglu@metu.edu.tr}
\affiliation{Department of Physics, Faculty of Arts and  Sciences,\\
             Middle East Technical University, 06531, Ankara, Turkey}

\date{\today}

\begin{abstract}
It is shown explicitly that when the characteristic vector field that
defines a G{\"o}del-type metric is also a Killing vector, there 
always exist closed timelike or null curves in spacetimes
described by such a metric. For these geometries, the geodesic curves 
are also shown to be characterized by a lower dimensional Lorentz 
force equation for a charged point particle in the relevant Riemannian 
background. Moreover, two explicit examples are given for which timelike 
and null geodesics can never be closed.
\end{abstract}


\maketitle

\section{\label{intro} Introduction}

The celebrated G{\"o}del metric \cite{god} solves the Einstein 
field equations with a homogeneous perfect fluid source and has a 
$G_{5}$ maximal symmetry \cite{kram}. The G{\"o}del spacetime admits 
closed timelike and closed null curves but contains no closed timelike
or closed null geodesics \cite{eh}. Moreover, the G{\"o}del universe 
is geodesically complete and it has neither a singularity nor a 
horizon.

In its original form, the G{\"o}del metric reads
\begin{equation}
ds^2 = -(dx^{0})^{2} + (dx^{1})^{2} - \frac{1}{2} \, e^{2 x^{1}} \, 
(dx^{2})^{2} + (dx^{3})^{2} - 2 \, e^{x^{1}} \, dx^{0} \, dx^{2} \, .
\label{godelm}
\end{equation}
[In fact G{\"o}del's original metric has an overall constant factor 
$a^{2}$ multiplying the line element. We have taken $a=1$ here.] 
A simple rearrangement of the terms yields that (\ref{godelm}) can 
also be written as
\begin{equation} 
ds^2 = (dx^{1})^{2} + \frac{1}{2} \, e^{2 x^{1}} \, 
(dx^{2})^{2} + (dx^{3})^{2} - (dx^{0} + e^{x^{1}} \, dx^{2})^{2} \, .
\label{god1}
\end{equation}
This form (\ref{god1}) suggests that the G{\"o}del metric can
be thought of as cast in the form
\begin{equation}
g_{\mu \nu}= h_{\mu \nu} - u_{\mu} \, u_{\nu} \; , \label{met}
\end{equation}
where the `background' $h_{\mu\nu}$ is a non-flat 3-metric and
\( u_{\mu} =  \delta_{\mu}^{0} + e^{x^{1}} \, \delta_{\mu}^{2} \)
is a timelike unit vector: \( g^{\mu\nu} \, u_{\mu} \, u_{\nu} = -1 \).
Notice, however, that (\ref{god1}) is not the only way of rearranging 
the terms in (\ref{godelm}). One could as well rewrite it as
\begin{equation}
ds^2 = (dx^{0})^{2} + (dx^{1})^{2} + (dx^{3})^{2} 
- \left( \sqrt{2} \, dx^{0} + \frac{1}{\sqrt{2}} \, 
e^{x^{1}} \, dx^{2} \right)^{2} \, , \label{god2}
\end{equation}
which again can be viewed in the form of (\ref{met}); however, this time
the `background' \( h_{\mu\nu}=\mbox{diag} \, (1, 1, 0, 1) \) describes
an obviously flat 3-dimensional spacetime and the new
\( u_{\mu} = \sqrt{2} \, \delta^{0}_{\mu} + (1/\sqrt{2}) \, e^{x^{1}} \, 
\delta^{2}_{\mu} \) is again a timelike unit vector in this new form. 

Inspired by these observations, a class of $D$-dimensional metrics 
were introduced, which were christened as {\it G{\"o}del-type metrics}, 
and used for producing new solutions to various
gravitational theories in diverse dimensions \cite{gks, gs}. Stated
briefly, these metrics are of the form (\ref{met}), where 
the background $h_{\mu \nu}$ is the metric of an Einstein space 
of a $(D-1)$-dimensional Riemannian geometry in the most general case
and $u^{\mu}$ is a timelike unit vector. A further assumption is that
both $h_{\mu \nu}$ and $u_{\mu}$ are independent of the fixed special 
coordinate $x^{k}$ with $0 \le k \le D-1$ and, moreover, that 
$h_{k\mu}=0$. A detailed analysis was given corresponding to the two 
distinct cases $u_{k}=$ constant and $u_{k} \neq $ constant, where 
$u^{\mu}=-\delta^{\mu}_{\;k}/u_{k}$, in \cite{gks} and \cite{gs}, 
respectively. It was shown that G{\"o}del-type metrics with $u_{k} =$ 
constant can be used in constructing solutions to the Einstein-Maxwell 
field equations with a dust distribution in $D$ dimensions and in 
that case the only essential field equation turns out to be the 
source-free `Maxwell's equation' in the relevant background \cite{gks}. 
One also finds that $u^{\mu}$ is tangent to a timelike geodesic curve
that is also a timelike Killing vector \cite{gks}. Moreover, it was 
established in \cite{gs} that when $u_{k} \neq $ constant, the 
conformally transformed G\"{o}del-type metrics can be used in 
solving a rather general class of Einstein-Maxwell-dilaton-3-form 
field theories in $D \geq 6$ dimensions and that, in this case, all 
field equations can be reduced to a simple `Maxwell equation' in 
the corresponding $(D-1)$-dimensional Riemannian background. 
However, with $u_{k} \neq $ constant, $u^{\mu}$ is no longer a
Killing vector field unlike the constant case \cite{gs}.
In these works, it was also shown that G\"{o}del-type metrics can 
be used in obtaining exact solutions to various supergravity theories, 
in which case $u_{k}$ may be considered as related to a dilaton 
field $\phi$ via the relationship $\phi = \ln{|u_{k}|}$. (See 
\cite{gks,gs} for the details.)

The existence of the closed timelike and closed null curves in the
G\"{o}del spacetime can be best inferred by transforming (\ref{godelm})
to the cylindrical coordinates in which case it reads
\begin{equation}
ds^2 = - d\tau^2 + dr^2 + dz^2 - \sinh^{2}{r} \, (\sinh^{2}{r} -1) 
\, d\varphi^2 + 2 \sqrt{2} \, \sinh^{2}{r} \, d\varphi \, d\tau 
\, . \label{godcy}
\end{equation}
It readily follows that the curve 
\( C = \{ (t, r, \varphi, z) \, | \, t=t_{0}, r = r_{0}, z = z_{0},
\varphi \in [0, 2\pi] \}, \) where $t_{0}, r_{0}$ and $z_{0}$ are
constants, is a closed timelike curve for \( r_{0} > \ln{(1+\sqrt{2})} \)
and a closed null curve for \( r_{0} = \ln{(1+\sqrt{2})} \).

There have been attempts to remove such closed timelike curves from
spacetimes described by G\"{o}del-type metrics by introducing observer 
dependent holographic screens \cite{bghv,bdebr} in the context of 
supergravity theories. Other `remedies' that have been put forward to
remove such curves involve the addition of scalar fields \cite{reti}, 
more specifically dilaton and axion fields \cite{bada}, to change 
the matter content, and to consider theories that involve higher order
curvature terms in their local gravity action \cite{acci,bar2}. The discussion
of the closed timelike curves in these works, e.g. \cite{bada, reti}, seems 
to be restricted to an investigation of the curves parametrized as the 
curve $C$ above, and thus solely on the general behaviour of the metric
component $g_{\varphi\varphi}$ since they use a general metric ansatz 
\[ ds^2 = -[dt + C(r) \, d\varphi]^{2} + D^{2}(r) \, d\varphi^{2} 
+ dr^2 + dz^2 \]
that respects cylindrical symmetry and generalizes the G{\"o}del metric
(\ref{godcy}). 

The discussion regarding the existence of closed timelike or null curves
and the behavior of geodesics in G\"{o}del-type metrics with 
$u_{k} = $ constant had to be kept concise in \cite{gks}, thus only
curves parametrized as the curve $C$ above were considered. However, it
is obvious that there can be other classes of curves that can be
both closed and timelike (or null). The aim of the present work is to 
provide a much more detailed analysis of these special curves in geometries 
described by such metrics. Specifically, it is going to be proved explicitly 
that {\em the non-flat spacetimes described by G\"{o}del-type metrics with both 
flat and non-flat backgrounds always have closed timelike or null curves}. 
As a separate discussion, it will also be shown that the geodesics of 
G\"{o}del-type metrics with constant $u_{k}$ are characterized by 
the $(D-1)$-dimensional Lorentz force equation for a charged point 
particle formulated in the corresponding Riemannian background.

\section{\label{ctcsimple} Closed timelike curves in G\"{o}del-type metrics
with flat backgrounds and constant $u_{k}$}

In this section, we will assume without loss of generality that the 
fixed special coordinate $x^{k}$ equals $x^{0} \equiv t$, the background 
$h_{\mu \nu}$ describes a flat Riemannian geometry, $h_{0\mu}=0$ 
and $u_{0}=1$. We will also take $D=4$ for simplicity but what follows 
can easily be generalized to higher dimensions.

In subsection 2.4 of \cite{gks}, spacetimes with the line element
\begin{equation}
ds^2 = d\rho^{2} + \rho^{2} d\phi^{2} + dz^2
- (dt + s(\rho,\phi) \, dz)^2 \, . \label{omet}
\end{equation}
were considered. It was found that this metric (\ref{omet}) is a 
solution of the charged dust field equations in four dimensions
provided $s(\rho,\phi)$ is a harmonic function in two dimensions.
Then the simplest possible choice, namely $s =$ constant and 
a rather plain and specific curve of the form 
\( \bar{C}=\{(t,\rho,\phi,z) \, | \, t=t_{0}, \rho=\rho_{0}, \phi=\phi_{0},
z \in [0, 2\pi) \},\) where $t_{0}, \rho_{0}$ and $\phi_{0}$ are constants,
was considered. Then it was shown that the tangent vector 
\( \bar{v}^{\mu}=(\partial/\partial z)^{\mu} \) is timelike if
\( (s(\rho_0,\phi_0))^2 > 1 \). However, it was argued that one
could exclude these closed timelike curves by considering the $z$ 
coordinate to be in the universal covering of this patch, namely by
taking $z$ to be on the real line $\mathbb{R}$. Finally, this
subsection was closed with the statement: {\it We thus conclude 
that the solutions we present correspond to spacetimes that contain 
both closed timelike and null curves and that contain neither of these 
depending on how one solves (7).} [Here (7) refers to the flat 3-dimensional
Euclidean source-free `Maxwell's equation' \( \partial_{i} \, f_{ij} =0 \),
where $i,j$ indices range from 1 to 3 and 
\( f_{ij} \equiv \partial_{i} u_{j} - \partial_{j} u_{i}\).] 

It readily follows that the `Maxwell equation' 
\( \partial_{i} \, f_{ij} =0 \) is equivalent to the Laplace equation 
for the (nontrivial) function $s(\rho,\phi)$ in two dimensions. In 
\cite{gks}, it was implicitly assumed that there were nontrivial 
harmonic functions suitable for the discussion carried on and the 
conclusion reached in subsection 2.4. However, we want to show in 
what follows that the previous discussion does not necessarily exclude 
the possibility of other closed timelike curves.

Let us, first of all, consider any smooth timelike curve $\Gamma$ in the
spacetime described by (\ref{omet}) with constant $\rho$ and $\phi$, so
that $(t(\eta), \rho_{0}, \phi_{0}, z(\eta))$ is a parametrization of
$\Gamma$ with a normalized tangent vector 
\begin{equation}
\left( \frac{d z}{d \eta} \right)^{2} - \left( \frac{d t}{d \eta} 
+ s_{0} \, \frac{d z}{d \eta} \right)^{2} = -1 \,, \qquad \mbox{where} \;\;
s_{0} \equiv s(\rho_{0}, \phi_{0}) = \mbox{constant} \, .
\label{gameqn}
\end{equation}
This may be solved for $dt/d\eta$ and formally integrated on $\eta$ to 
get
\begin{equation}
t(\eta) = t(0) - s_{0} \, [z(\eta) - z(0)] + \epsilon \,
\int_{0}^{\eta} \left[ 1+ \left( \frac{d z}{d \eta^{\prime}} \right)^{2} 
\right]^{1/2} \, d\eta^{\prime} \,, \qquad \mbox{where} \;\; 
\epsilon = \pm 1 \, . \label{tetaeq} 
\end{equation}
Since the integral in (\ref{tetaeq}) is an increasing function of $\eta$,
this shows that there are no smooth timelike curves along which both $t$
and $z$ simultaneously recover their initial values and, therefore, there
are no closed timelike curves of this type for the metric (\ref{omet}). 
However, for $\eta \gg 1$, one may have timelike curves where both $t$
and $z$ perform rather large excursions, but then come back to values that
are close to their initial values. 

This brings to mind the question whether it could be possible to obtain
closed timelike curves if one allows also a variation in $\rho$ and/or
$\phi$. As an example, consistent with the assumption of being harmonic,
let us assume that in a neighborhood of a point $(\rho_{0}, \phi_{0})$,
for fixed $\phi$, and take the function $s(\rho,\phi)$ to be of the
form
\begin{equation}
s(\rho,\phi) = s_{0} + s_{1} \, (\rho - \rho_{0}) \, , 
\end{equation}
where $s_{0}$ and $s_{1}$ are constants. Then, one can check that the
curve
\begin{eqnarray}
\rho(\eta) & = & \rho_{0} + a \, \cos{\eta} \,, \quad 
\phi = \phi_{0} \,, \quad z(\eta) = a \, \sin{\eta} \,, \nonumber \\
t(\eta) & = & - a \, \left[ s_{0} + \frac{a s_{1}}{2} \, \cos{\eta} \right]
\, \sin{\eta} \label{acurve}
\end{eqnarray}
is a smooth timelike curve with normalized tangent vector, provided that
$a$ is a solution of
\begin{equation}
s_{1}^{2} \, a^{4} - 4 a^{2} - 4 = 0 \, . \label{aeqn}
\end{equation}
A suitable solution of (\ref{aeqn}) is 
\begin{equation}
a = \frac{\sqrt{2+2 \sqrt{1+s_{1}^{2}}}}{s_{1}} \, , \qquad
s_{1} \neq 0 \; ,\label{asoln}
\end{equation}
which is well defined for all values of $s_{1}$. Since the coordinates
$(t, \rho, z)$ are all periodic in $\eta$, one concludes that (\ref{acurve}) 
plus (\ref{asoln}) represent a smooth closed timelike curve for the
metric (\ref{omet}). 

This was only a counter example for the assertion that (\ref{omet})
is an example of a spacetime without closed timelike curves. It follows
from this analysis that in order to determine the absence (or the presence)
of closed timelike curves, one would need to consider in each case the
explicit form of the function $s(\rho,\phi)$. As for another, more general,
example, consider
\begin{equation}
s_{n}(\rho,\phi) = \gamma + \rho^{n} \, 
(\alpha \, \cos{n \phi} + \beta \, \sin{n \phi}) \, , 
\quad n \in \mathbb{Z}^{+} \; , \label{esen}
\end{equation}
where $\alpha$, $\beta$ and $\gamma$ are real constants. [Obviously this
function is related to the real and imaginary parts of the complex analytic
function \( g(\zeta) = \zeta^{n} \,, n \in \mathbb{Z}^{+} .\)] It readily
follows that the nontrivial components of the Maxwell (matter) field 
\[ f_{\rho z} = n \, \rho^{n-1} \, 
(\alpha \, \cos{n \phi} + \beta \, \sin{n \phi}) \; , \quad
f_{\phi z} = n \, \rho^{n} \, 
(-\alpha \, \sin{n \phi} + \beta \, \cos{n \phi}) \]
now depend on the constants $\alpha$ and $\beta$, and do not vanish for all
$(\rho,\phi)$ unless \( \alpha, \beta \rightarrow 0 .\)

Let us now take the curve $\tilde{\Gamma}$ (similar to the curve $\Gamma$
above) defined by $(t(\eta),\rho(\eta),\phi_{0},z(\eta))$, where one can
assume that the parameter $\eta$ takes values in $[0, 2 \pi]$ without
any loss of generality,
\begin{equation}
\rho(\eta) = \rho_{0} + a \cos{\eta} \, , \quad z(\eta) = a \sin{\eta} \, ,
\end{equation}
and $\phi_{0}$, $\rho_{0}$ and $a$ are real constants as before. Using 
the line element (\ref{omet}) with $s_{n}(\rho,\phi)$ given in (\ref{esen}) 
and solving the constraint equation that normalizes the tangent vector 
of this curve $\tilde{\Gamma}$ to unity, one finds that
\begin{equation}
\frac{dt}{d\eta} = \epsilon \sqrt{1+a^2} - a \, s_{n}(\rho,\phi_{0}) \,
\cos{\eta} \,, \qquad \mbox{where, again,} \;\; 
\epsilon = \pm 1 \, , \label{tdenk}
\end{equation}
and
\begin{equation}
s_{n}(\rho,\phi_{0}) = \gamma + \sigma_{n} (\rho_{0} + a \cos{\eta})^{n}
\quad \mbox{with} \;\; 
\sigma_{n} \equiv \alpha \, \cos{n \phi_{0}} + \beta \, \sin{n \phi_{0}} \, .
\end{equation}
When (\ref{tdenk}) is integrated for $t$, the outcome can be put in a form
\begin{equation}
t(\eta) = f_{n}(\eta) + (A_{n} \, \sigma_{n} + \epsilon \sqrt{1+a^2}) \eta
+ t_{0} \; , \quad t_{0} = \mbox{integration constant} \, , \label{tcoz}
\end{equation}
where $f_{n}(\eta)$ is a periodic function of $\eta$ that consists of a 
linear sum of $\sin{m \eta}$ terms (with $m=0,1,\dots, n+1$) and $A_{n}$ is
another constant term, both of which depend naturally on the choice of the
constant $n \in \mathbb{Z}^{+}$. [Using the notation introduced thus far, 
one gets
\[ A_{n} = \left\{
\begin{array}{lc}
-a^2/2 \, , & n = 1 \\
-a^2 \, \rho_{0} \, , & n = 2
\end{array} \right. \; , \]
\[ f_{n}(\eta) = \left\{
\begin{array}{lc}
-a (\gamma + \sigma_{1} \, \rho_{0}) \sin{\eta} - 
(a^2 \, \sigma_{1}/4) \, \sin{2 \eta} \, , & n = 1 \\
-( a(\gamma + \sigma_{2} \rho_{0}^{2}) + 3 a^{3} \sigma_{2}/4 )
\sin{\eta} - (a^{2} \rho_{0} \sigma_{2}/2) \sin{2 \eta} 
- (a^{3} \sigma_{2}/12) \sin{3 \eta} \, , & n = 2
\end{array} \, , \right. \]
for the simplest cases $n=1$ and $n=2$.] Similar to how the $t(\eta)$
part of the curve $\Gamma$ was made a periodic function of $\eta$ above,
one can set the coefficient of the $\eta$ term on the right hand side 
of (\ref{tcoz}) to zero by suitably choosing the constant $A_{n}$, provided
the constant $\sigma_{n}$ does not tend to zero or that $s_{n}(\rho,\phi)$ 
does not go to a constant. One again concludes that $\tilde{\Gamma}$, 
as described above, represents a smooth closed timelike curve for the
metric (\ref{omet}). 

A natural question to address then is whether closed timelike or null curves 
exist for all harmonic functions. The answer one finds is as follows: 

Consider the most general curve $C$ defined by 
$(t(\eta),\rho(\eta),\phi(\eta),z(\eta))$, where
the arc-length parameter $\eta$ takes values in the simple (generic) 
interval $[0, 2\pi]$. Normalizing the tangent vector of this curve
$C$ to unity in the geometry described by (\ref{omet}), one finds
\begin{equation}
\frac{dt}{d\eta} = - s(\rho,\phi) \, \frac{dz}{d\eta} + \epsilon
\sqrt{\lambda + \left( \frac{d\rho}{d\eta} \right)^2 + 
\left( \frac{dz}{d\eta} \right)^2 + \rho^{2} \,
\left( \frac{d\phi}{d\eta} \right)^2}  \,,\label{teq}
\end{equation}
where $\lambda = 0$ for null and $\lambda = 1$ for timelike curves. 
Now let the parametrizations of $\rho$, $\phi$ and $z$ be all periodic 
functions in $\eta$. Then the terms in the square root in (\ref{teq}) 
can be expanded in a Fourier series in the interval $[0,2\pi]$ and it 
is clear that this Fourier series expansion has a non-negative constant 
term in it which looks like \footnote{Strictly speaking, $B$ is 
positive definite for $\lambda = 1$ and non-negative for $\lambda = 0$.
When $\lambda = 0$, $B=0$ iff $\rho$, $\phi$ and $z$ are all constants,
i.e. $s(\rho,\phi) =$ constant which we have already discussed. Hence,
for the discussion that follows we assume that $B>0$.}
\[ B = \frac{1}{2\pi} \, \int_{0}^{2\pi} \,
\sqrt{\lambda + \left( \frac{d\rho}{d\eta} \right)^2 + 
\left( \frac{dz}{d\eta} \right)^2 + \rho^{2} \,
\left( \frac{d\phi}{d\eta} \right)^2} \; d\eta \, > 0 \, . \]
As it stands, since $B \neq 0$, $t(\eta)$ naturally picks up a non-periodic 
piece $B \eta$ from the second term on the right hand side of (\ref{teq})
and in order to have no closed timelike (or null) curves, it must 
be that
\begin{equation}
\int_{0}^{2\pi} \, s(\rho(\eta),\phi(\eta)) \, \frac{dz}{d\eta} \, 
d\eta = 0 \label{sint} 
\end{equation}
for all arbitrary periodic functions $z(\eta)$. However, since $\rho(\eta)$
and $\phi(\eta)$ are periodic functions of $\eta$, it readily follows that
$s(\rho(\eta),\phi(\eta))$ is also periodic in $\eta$. [In fact, it is
enough to demand that $s(\rho,\phi)$ is a continuous function, rather 
than a harmonic function, of its arguments for the discussion that
follows. We would like to stress that the condition for being harmonic 
followed from the Maxwell equation (7) of \cite{gks}.] In this case, one
can expand $s(\rho(\eta),\phi(\eta))$ in a Fourier series in $\eta$ as
\[ s(\rho(\eta),\phi(\eta)) = a_{0} + g(\eta) =
a_{0} + \sum_{p=1}^{\infty} \, (a_{p} \, \cos{p\eta} 
+ b_{p} \, \sin{p\eta} ) \,  \]
where $a_{0}$, $a_{k}$ and $b_{k}$ are the Fourier coefficients in the
usual manner. Now (\ref{sint}) implies that
\[ \int_{0}^{2\pi} \, g(\eta) \, \frac{dz}{d\eta} \, d\eta = 0 \, . \]
If one chooses the periodic function $z(\eta)$ so that 
$dz/d\eta = g(\eta)$, then
\[ \int_{0}^{2\pi} \, (g(\eta))^{2} \, d\eta = 0 \,, \]
which is possible only if $g(\eta)=0$, and hence $s(\rho,\phi)=$ constant.
Therefore, unless $s(\rho,\phi)=$ constant, one can always cancel out 
the contribution of the $B$ term above and find a closed timelike 
or null curve in the spacetime described by (\ref{met}).

We thus see that the conclusion reached at the end of subsection 2.4 of
\cite{gks} needs to be corrected as follows: There exists no nontrivial 
(non-constant) harmonic function $s(\rho,\phi)$ such that the spacetime 
described by (\ref{omet}) has no closed timelike or null curves; i.e. 
{\em one can always find a closed timelike or null curve in the geometry 
of (\ref{omet}) given an arbitrary harmonic function $s(\rho,\phi)$.}

\section{\label{ctchard} Closed timelike curves in G\"{o}del-type metrics
with non-flat backgrounds and constant $u_{k}$}

Let us now see how the results of the section \ref{ctcsimple} can also
be applied to G\"{o}del-type metrics with non-flat backgrounds but
with constant $u_{k}$. Let us once again assume without loss of generality
that the fixed special coordinate $x^{k}$ equals $x^{0} \equiv t$, the 
background $h_{\mu \nu}$ is the metric of an Einstein space of a 
$(D-1)$-dimensional (non-flat) Riemannian geometry, $h_{0\mu}=0$ 
and $u_{0}=1$. To keep the discussion simple, let us take $D=4$ as before;
however, the following can be generalized to higher dimensions without any
difficulty.

In \cite{gks}, it was found that the G\"{o}del-type metric
\begin{equation}
ds^2 = h_{ij}(x^{\ell}) \, dx^{i} \, dx^{j} - 
(dt + u_{i}(x^{\ell}) \, dx^{i})^{2} \, , \label{hmet}
\end{equation}
where indices $i, j, \ell$ range from 1 to 3, is a solution to 
the Einstein-Maxwell field equations with a dust distribution provided
that the 3-dimensional source-free `Maxwell equation'
\begin{equation}
\partial_{\alpha} (\bar{h}^{\alpha\mu} \, \bar{h}^{\beta\nu} \, \sqrt{|h|} \,
f_{\mu\nu}) = 0 \, , \label{max}
\end{equation}
where \( f_{\mu\nu} = \partial_{\mu} u_{\nu} - \partial_{\nu} u_{\mu} \)
as before and $\bar{h}^{\mu\nu}$ is the 3-dimensional inverse of
$h_{\mu\nu}$, is satisfied. Starting with a specific background 
$h_{\mu\nu}$, one should be able, in principle, to solve (\ref{max}) 
and use the solution vector $u_{i}$ to write down (\ref{hmet}) 
explicitly. It should be kept in mind that $h_{ij} = h_{ij}(x^{\ell})$ 
by assumption and, solving (\ref{max}), $u_{i} = u_{i}(x^{\ell})$ in 
the most general case.

Now let us consider a general curve $C$ defined by 
$(t(\eta), x^{i}(\eta))$ parametrized with the arc-length parameter 
$\eta \in [0, 2\pi]$ as in section \ref{ctcsimple}. Normalizing the
tangent vector of $C$ to unity by using (\ref{hmet}), one obtains
\begin{equation}
\frac{dt}{d\eta} = - u_{i}(x^{\ell}) \, \frac{dx^{i}}{d\eta} + \epsilon
\sqrt{\lambda + h_{ij}(x^{\ell}) \, \frac{dx^{i}}{d\eta} 
\, \frac{dx^{j}}{d\eta}}  \,,\label{newteq}
\end{equation}
where $\epsilon = \pm 1$, $\lambda = 0$ for null and $\lambda = 1$ 
for timelike curves. Now let the parametrization of each $x^{i}$ be 
a periodic function in $\eta$. One can then follow similar steps to the
ones described in section \ref{ctcsimple} and use Fourier series
expansion of the terms on the right hand side of (\ref{newteq}). One
finds that the Fourier series expansion has a non-vanishing constant
term of the form \footnote{Once again, $\beta$ is in fact positive 
definite for $\lambda = 1$ and non-negative for $\lambda = 0$.
However, when $\lambda = 0$, $\beta=0$ iff each $x^{i}(\eta)$ is a constant,
and in that case (\ref{hmet}) obviously has no closed null curves
since by assumption the background $h_{\mu \nu}$ has positive definite
signature. Hence, we assume that $\beta>0$ for what follows.}
\[ \beta = \frac{1}{2\pi} \, \int_{0}^{2\pi} \,
\sqrt{\lambda + h_{ij}(x^{\ell}) \, \frac{dx^{i}}{d\eta} 
\, \frac{dx^{j}}{d\eta}} \; d\eta \, > 0 \, , \]
which contributes a term $\beta \eta$ to the integration of (\ref{newteq})
for $t(\eta)$. 

In order to have no closed timelike (or null) curves, one must have
\begin{equation}
\sum_{i=1}^{3} \, \int_{0}^{2\pi} \, u_{i}(x^{\ell}) \, 
\frac{dx^{i}}{d\eta} \, d\eta = 0 \label{cond} 
\end{equation}
for all arbitrary periodic functions $x^{i}(\eta)$. However, each
$u_{i}(x^{\ell})$ is also periodic in $\eta$ since each $x^{\ell}(\eta)$
is. Moreover, even though their explicit form may not be available,
the functions $u_{i}(x^{\ell})$ are also continuous functions of $\eta$
since they satisfy a second order (nonlinear) partial differential
equation (\ref{max}). Thus these functions may also be expanded in a
Fourier series as
\[ u_{i}(x^{\ell}) = a_{i0} + g_{i}(\eta) =
a_{i0} + \sum_{p=1}^{\infty} \, (a_{ip} \, \cos{p\eta} 
+ b_{ip} \, \sin{p\eta} ) \,  \]
with $a_{i0}$, $a_{ip}$ and $b_{ip}$ denoting the Fourier coefficients 
in the usual sense. Now (\ref{cond}) implies that
\[ \sum_{i=1}^{3} \, \int_{0}^{2\pi} \, g_{i}(\eta) \, 
\frac{dx^{i}}{d\eta} \, d\eta = 0 \, . \]
If one chooses the periodic functions $x^{i}(\eta)$ such that 
$dx^{i}/d\eta = g_{i}(\eta)$ for each $i$, then
\[ \sum_{i=1}^{3} \, \int_{0}^{2\pi} \, (g_{i}(\eta))^{2} \, d\eta = 0 \,, \]
which is possible only if $g_{i}(\eta)=0$ and hence $u_{i}(x^{\ell}) =$ 
constant for each $i$. However, one can then define a new coordinate
$d\tau = dt + u_{i} \, dx^{i} $ and can immediately see that (\ref{hmet})
has no closed timelike or null curves. Hence, unless each one of
$u_{i}(x^{\ell})$ is a constant, one can always find a closed timelike 
or null curve in the spacetime (\ref{hmet}) by smartly choosing a
parametrization that cancels out the contribution of the $\beta$ term
in the $t(\eta)$ equation.

We thus see that the conclusion reached at the end of section 
\ref{ctcsimple} can be extended to include all G\"{o}del-type 
metrics of the form (\ref{hmet}). When each $u_{i}(x^{\ell}) =$ 
constant, the spacetime is obviously flat and there exists no 
closed timelike or null curves. In retrospect, we have proved that 
{\em the spacetimes described by G\"{o}del-type metrics with both flat and 
non-flat backgrounds always have closed timelike or null curves,
provided that at least one of the $u_{i}(x^{\ell}) \neq$ constant}.

\section{\label{geod} Geodesics of G\"{o}del-type metrics with 
constant $u_{k}$}

Let us now examine the geodesics of G\"{o}del-type metrics. In \cite{gks},
it was shown that G\"{o}del-type metrics with flat backgrounds and
constant $u_{k}$ have their geodesics described by the Lorentz force 
equation for a charged particle in the corresponding $(D-1)$-dimensional 
Riemannian background. Let us now look into how the geodesics of a 
G\"{o}del-type metric with a general non-flat background (but still with
constant $u_{k}$) behave. To simplify the discussion, let us again keep
the general assumptions listed at the beginning of section \ref{ctchard},
but this time let us keep $D$ arbitrary.

The inverse of the G\"{o}del-type metric (\ref{met}) is
(as already given in (2) of \cite{gks})
\begin{equation}
g^{\mu \nu} = \bar{h}^{\mu\nu} + (-1+\bar{h}^{\alpha\beta} \, u_{\alpha} \, 
u_{\beta}) \, u^{\mu} \, u^{\nu} 
+ u^{\mu} (\bar{h}^{\nu\alpha} \, u_{\alpha}\,) 
+ u^{\nu} (\bar{h}^{\mu\alpha} \, u_{\alpha}\,) \; , \label{invmet}
\end{equation}
with $\bar{h}^{\mu\nu}$ denoting the $(D-1)$-dimensional inverse of 
$h_{\mu \nu}$, as before. The Christoffel symbols of (\ref{met}) 
in this case were also given by (37) of \cite{gks} as
\begin{equation}
\Gamma^{\mu}\,_{\alpha \beta} = \gamma^{\mu}\,_{\alpha \beta}
+ \frac{1}{2}\, (u_{\alpha} \, f^{\mu}\,_{\beta}
+ u_{\beta} \, f^{\mu}\,_{\alpha}) - \frac{1}{2} \, u^{\mu} \,
(u_{\alpha \vert \beta} + u_{\beta \vert \alpha}) \, . \label{chris}
\end{equation}
in terms of the Christoffel symbols $\gamma^{\mu}\,_{\alpha \beta}$ 
of $h_{\mu\nu}$ and the `Maxwell field' 
\( f_{\mu\nu} \equiv \partial_{\mu} u_{\nu} - \partial_{\nu} u_{\mu} \),
as before. Here a vertical stroke denotes a covariant derivative with 
respect to $h_{\mu\nu}$ and it should be kept in mind that the indices
on $f$ and $u$ are raised and lowered by the metric $g_{\mu\nu}$.

With all the ingredients already at hand, let us now consider a geodesic
curve in the spacetime described by (\ref{met}) which is 
parametrized as $x^{\mu}(\tau)$.  Using (\ref{chris}) and denoting 
derivative with respect to the affine parameter $\tau$ by a dot, 
the geodesic equation yields
\[ \ddot{x}^{\mu} + \gamma^{\mu}\,_{\alpha \beta} \, \dot{x}^{\alpha} \,
\dot{x}^{\beta} + (u_{\alpha} \, \dot{x}^{\alpha}) \, 
( f^{\mu}\,_{\beta} \, \dot{x}^{\beta}) - u^{\mu} \,
( u_{\alpha \vert \beta} \, \dot{x}^{\alpha} \, \dot{x}^{\beta}) = 0 \, . \]
Noting that \( u_{\alpha, \, \beta} \, \dot{x}^{\beta} = \dot{u}_{\alpha} \),
writing $f^{\mu}\,_{\beta}$ explicitly via the inverse of the metric
(\ref{invmet}) and using \( u^{\alpha} \, f_{\mu \alpha} = 0 \), this 
becomes
\begin{equation}
\ddot{x}^{\mu} + \gamma^{\mu}\,_{\alpha \beta} \, \dot{x}^{\alpha} \,
\dot{x}^{\beta} + u_{\alpha} \, \dot{x}^{\alpha} \, 
( \bar{h}^{\mu\sigma} + u^{\mu} \, \bar{h}^{\sigma\nu} \, u_{\nu} ) \,
f_{\sigma\beta} \, \dot{x}^{\beta} - u^{\mu} \, 
( \dot{u}_{\alpha} \, \dot{x}^{\alpha} - \gamma^{\sigma}\,_{\alpha \beta} 
\, u_{\sigma}  \, \dot{x}^{\alpha} \, \dot{x}^{\beta} ) = 0 \, . \label{geod1}
\end{equation}
Now contracting this with $u_{\mu}$ and using \( u_{\mu} u^{\mu} = -1 \),
one obtains a constant of motion for the geodesic equation as
\begin{equation} 
u_{\mu} \, \dot{x}^{\mu} = - e = \; \mbox{constant} \, . \label{geodch}
\end{equation}
Meanwhile setting the free index $\mu=i$ in (\ref{geod1}) and using
\( u^{\mu} = - \delta^{\mu}_{\;0} \), one also finds
\[ \ddot{x}^{i} + \gamma^{i}\,_{\alpha \beta} \, \dot{x}^{\alpha} \,
\dot{x}^{\beta} - e \, (\bar{h}^{i\sigma} \, f_{\sigma\beta} \, 
\dot{x}^{\beta}) = 0 \, , \] 
or equivalently
\begin{equation}
\dot{x}^{\mu} \, \dot{x}^{i}_{\; \vert \mu}  = e \, \bar{h}^{i\sigma} 
\, f_{\sigma\mu} \, \dot{x}^{\mu} \, , \;\;\;\; (i=1, 2, \dots, D-1) \,,
\label{geod2}
\end{equation}
i.e. the analogous $(D-1)$-dimensional Lorentz force equation for a charged
point particle (of charge/mass ratio equal to $e$) written in the 
corresponding Riemannian background. Moreover, contracting (\ref{geod2}) 
further by $h_{ij} \, \dot{x}^{j}$ and using the antisymmetry of 
$f_{\mu\nu}$, one obtains a second constant of motion
\[ h_{ij} \, \dot{x}^{i} \, \dot{x}^{j} = \ell^{2} > 0 \, . \]
Since 
\begin{equation}
g_{\mu\nu} \, \dot{x}^{\mu} \, \dot{x}^{\nu} = 
h_{ij} \, \dot{x}^{i} \, \dot{x}^{j} - (u_{\mu} \, \dot{x}^{\mu})^{2}
= \ell^{2} - e^2 \, , \label{normeq}
\end{equation}
one concludes that the nature of the geodesics necessarily depends on the
sign of $\ell^{2}-e^2$.

Note in fact that the existence of these two constants of motion
$e$ and $\ell$ follow naturally from the very structure of the metric
(\ref{met}). (\ref{geodch}) is a simple consequence of the fact that
$u^{\mu}$ is a Killing vector and tangent to a timelike geodesic curve
when $u_{k} =$ constant, as already pointed out in section \ref{intro}.
Furthermore, since \( g_{\mu\nu} \, \dot{x}^{\mu} \, \dot{x}^{\nu} = -1, 0 \)
along the geodesic curve, (\ref{normeq}) follows immediately. However
one still needs (\ref{invmet}) and (\ref{chris}) to obtain (\ref{geod2})
explicitly.

Let us now take a specific example and examine the behavior of the geodesics
of this metric in more detail. Consider the simple solution discussed in
section 2.1 of \cite{gks}. These odd dimensional G\"{o}del-type metrics with 
flat backgrounds are simply characterized by 
\( u_{i} = b \, J_{ij} \, x^{j} / 2, \) where  $b$ is a real constant and
$J_{ij}$ is fully antisymmetric with constant components that satisfy
\( J^{k}\,_{j} \, J^{i}\,_{k} = - \delta^{i}_{j} \, . \) These yield
$f_{ij} = b J_{ij}$, $f_{0 \mu}=0$ and it is readily seen that the `Maxwell
equation'  \( \partial_{i} \, f_{ij} =0 \) holds trivially. It was shown
in \cite{gks} that this solution can be thought of as describing a
spacetime filled with dust or as a solution to the Einstein-Maxwell theory
when $D=5$. However, for odd $D>5$, it could as well be considered as a
solution of Einstein theory coupled with a perfect fluid source with 
negative pressure.

Employing these specifications, the $(D-1)$-dimensional Lorentz force 
equation (\ref{geod2}) can be put in the form 
\begin{equation} 
\ddot{{\bf x}} = \omega \, {\bf J} \, \dot{{\bf x}} \, , \label{xddot}
\end{equation}
where we have defined $\omega = eb$ and used matrix notation so that
${\bf x}$ denotes the $(D-1) \times 1$ column vector with entries
$x^{i}$ and ${\bf J}$ denotes the $(D-1) \times (D-1)$ matrix with
components $J_{ij}$. Defining a second column vector ${\bf y}$ as 
${\bf y} = \dot{{\bf x}}$, (\ref{xddot}) can be integrated as
\[ {\bf y} = e^{\omega \tau {\bf J}} \, {\bf x_{0}} \, , \]
for some constant column vector ${\bf x_{0}}$. Since 
${\bf J}^{2} = - {\bf I}_{D-1}$ (the $(D-1) \times (D-1)$ identity
matrix and keep in mind that $D$ is odd) by construction, this can also 
be written as
\[ {\bf y} = \dot{{\bf x}} = (\cos{\omega\tau} + {\bf J} \, 
\sin{\omega\tau}) \, {\bf x_{0}} \, , \]
or integrating once again
\begin{equation} 
{\bf x}(\tau) = \frac{1}{\omega} \, (\sin{\omega\tau} - {\bf J} \, 
\cos{\omega\tau}) \, {\bf x_{0}} + {\bf x_{1}} \label{xvsol}
\end{equation}
for a new constant column vector ${\bf x_{1}}$. Notice that
\[ ({\bf x} - {\bf x_{1}})^{T} \, ({\bf x} - {\bf x_{1}}) =
\frac{1}{\omega^{2}} \, {\bf x_{0}}^{T} \, {\bf x_{0}} \, . \]
Defining a scalar as 
${\cal R}^{2} \equiv {\bf x_{0}}^{T} \, {\bf x_{0}}/\omega^{2}$, one
finds that $|{\bf x} - {\bf x_{1}}|^{2} = {\cal R}^{2}$, which in
general is the equation of a $(D-2)$-dimensional sphere with radius
${\cal R}$ and center at ${\bf x_{1}}$.

The equation describing $x^{0}$ is given by (\ref{geodch}), which reads
\[ \dot{x}^{0} + \frac{b}{2} \, \dot{{\bf x}}^{T} \, 
{\bf J} \, {\bf x} + e = 0 \]
in matrix form. Using (\ref{xvsol}), this can be written as 
\[ \dot{x}^{0} + \left( e + \frac{1}{2} \, b \omega {\cal R}^{2} \right) +
\frac{b}{2} \, (b_{1} \, \sin{\omega\tau} + b_{2} \, \cos{\omega\tau}) = 0 \]
for some constants $b_{1} \equiv {\bf x_{0}}^{T} \, {\bf x_{1}}$ and
$b_{2} \equiv {\bf x_{0}}^{T} \, {\bf J} \, {\bf x_{1}}$. Integrating
this equation, one finally finds
\begin{equation}
x^{0}(\tau) = -e \, \left( 1 + \frac{1}{2} \, b^{2} {\cal R}^{2} \right) 
\, \tau + \frac{1}{2e} \, (b_{1} \, \cos{\omega\tau} - 
b_{2} \, \sin{\omega\tau}) + b_{3} \label{x0sol}
\end{equation}
for a new integration constant $b_{3}$. Note that the case for the null
geodesics corresponds simply to taking \( b {\cal R} = 1 \). Since the 
coefficient of the term linear in $\tau$ is never zero, $x^{0}(\tau)$ 
can not be periodic. Hence we conclude that the geodesic curves move 
on $\mathbb{R} \times S^{D-2}$ and are complete for this special solution 
in odd dimensions.

Note, in fact, that this result can also be generalized to the case 
when the rank of ${\bf J}$ is $2k$ where $2 \le 2k \le D-1$ and 
${\bf J}^{2} = - {\bf I}_{2k}$, the $2k \times 2k$ identity matrix. Then 
$2k$ of the spatial coordinates will obey the Lorentz force equation
\[ \ddot{x}^{i} = \omega \, J_{ij} \, \dot{x}^{j} \, , \qquad
(i = 1, 2, \dots, 2k) \, , \]
whereas the remaining spatial ones will simply satisfy
\[ \ddot{x}^{i} = 0 \, , \qquad (i = 2k+1, \dots, D-1) \, . \]
Therefore the $2k$-dimensional part will be periodic in $\tau$ and
will describe a $(2k-1)$-dimensional sphere whereas the remaining parts
will be linear in $\tau$. Hence the geodesic curves will move on 
$\mathbb{R}^{D-2k} \times S^{2k-1}$ in this case.

As yet another example, one can also consider the geodesics of the spacetime
described by (\ref{omet}). If one uses Cartesian coordinates $(x, y, z)$
instead of the circular cylindrical coordinates used in (\ref{omet}), the
geodesics can alternatively be found by extremizing the Lagrangian
\begin{equation}
L = \dot{x}^{2} + \dot{y}^{2} + \dot{z}^{2} - 
(\dot{t} + s(x(\tau), y(\tau)) \, \dot{z})^{2} = \lambda \, , \label{action}
\end{equation}
where $s(x, y)$ is a harmonic function of its arguments, dot denotes
differentiation with respect to the affine parameter $\tau$ and 
$\lambda = -1, 0$ for timelike and null geodesics, respectively. One
easily obtains two constants of motion from the $t$ and $z$ variations as
\begin{equation}
\dot{t} + s(x(\tau), y(\tau)) \, \dot{z} = -e = \mbox{constant}, \qquad
\mbox{and} \qquad 
\dot{z} + e \, s(x(\tau), y(\tau)) = \ell = \mbox{constant} \,, \label{tzeqn}
\end{equation}
whereas the $x$ and $y$ variations yield
\begin{equation}
\ddot{x} - e \, (\partial s/ \partial x) \, (\ell - e \, s(x, y)) = 0 
\qquad \mbox{and} \qquad 
\ddot{y} - e \, (\partial s/ \partial y) \, (\ell - e \, s(x, y)) = 0 \, , 
\label{xyeqn}
\end{equation}
respectively. Note that using the constants of motion in the expression
(\ref{action}) for the Lagrangian, one now also has the constraint
\begin{equation}
\dot{x}^{2} + \dot{y}^{2} = \lambda + e^{2} - 
(\ell - e \, s(x(\tau), y(\tau)))^{2} \, . \label{constr}
\end{equation}
In general, this system of coupled ordinary differential equations can
be solved given the explicit form of the harmonic function $s(x, y)$.

Let us now consider as a simple example, the harmonic function
\begin{equation}
s(x, y) = \gamma + \alpha \, x + \beta \, y \, , \label{sline}
\end{equation}
corresponding to the $n=1$ case in (\ref{esen}). Keeping in mind that
we are after closed timelike or null geodesics, one can now easily
integrate (\ref{xyeqn}), set the coefficient of the linear term
in $\tau$ to zero and obtain the periodic functions $x(\tau)$ and
$y(\tau)$. Using these in (\ref{tzeqn}), one then calculates $z(\tau)$
(which itself turns out to be a periodic function in $\tau$) and
$t(\tau)$ as
\[ t(\tau) = -\frac{e}{2} \, \left( 1 - \frac{\lambda}{e^2} \right) \tau
+ g(\tau) \, , \]
where the constraint (\ref{constr}) has been used and $g(\tau)$ contains
all the parts periodic in $\tau$. Hence for closed geodesics, one needs
that \( \lambda = e^{2} \)! However, this is not possible \footnote{Note 
that when $e=0$, (\ref{xyeqn}) and (\ref{tzeqn}) are easily solved yielding
$x(\tau)$, $y(\tau)$ and $z(\tau)$ as linear functions in $\tau$ and
$t(\tau)$ as a quadratic function in $\tau$, which obviously do not
describe closed geodesics then.} and one concludes that there are 
no closed timelike or null geodesics in the spacetime described 
by (\ref{omet}) when $s(x, y)$ is given by (\ref{sline}). We conjecture
that this result can also be generalized to the case of the more 
general harmonic function $s(x, y)$ given by (\ref{esen}).

\section{\label{conc} Conclusions}

In this work, it has been verified explicitly that {\em the spacetimes 
described by G\"{o}del-type metrics with both flat and non-flat 
backgrounds always have closed timelike or null curves, provided that
at least one of the $u_{i}(x^{\ell}) \neq$ constant}. It has also 
been shown that the geodesics of G\"{o}del-type 
metrics with constant $u_{k}$ are characterized by the 
$(D-1)$-dimensional Lorentz force equation for a charged point 
particle formulated in the corresponding Riemannian background.
A specific example in odd dimensions has been considered for which 
timelike and null geodesics are complete and never closed. It has
also been laid out in a separate example that there are no closed 
timelike or null geodesics in the spacetime described by (\ref{omet}) 
when $s(x, y)$ is a linear function of its arguments.

One possible direction to look at would be to examine the
existence of closed timelike curves in spacetimes described by
the most general G\"{o}del-type metrics with non-flat backgrounds
and non-constant $u_{k}$. 

\section{\label{acknow} Acknowledgments}

R.J.G. is supported by CONICET (Argentina) and in part by a  grant
from the National University of C\'{o}rdoba, (Argentina) and by grant 
NSF-INT-0204937 of the National Science Foundation of the USA.
M.G., A.K. and {\"O}.S. are partially supported by the Scientific 
and Technical Research Council of Turkey (T{\"U}B\.{I}TAK). M.G. is
also supported in part by the Turkish Academy of Sciences (T{\"U}BA).

\end{document}